# Quantum Mechanical Considerations on Algebraic Structure of Central Molecular Chirality

SALVATORE CAPOZZIELLO[a] AND ALESSANDRA LATTANZI[b]

[a]*Dipartimento di Fisica "E. R. Caianiello", INFN sez. di Napoli, Università di Salerno, Via S. Allende, 84081, Baronissi, Salerno, Italy*
[b]*Dipartimento di Chimica, Università di Salerno, Via S. Allende, 84081, Baronissi, Salerno, Italy*
E-mail: capozziello@sa.infn.it; lattanzi@unisa.it

*ABSTRACT*   The chiral algebra of tetrahedral molecules, derived from Fischer's projections, is discussed in the framework of quantum mechanics. A "quantum chiral algebra" is obtained whose operators, acting as rotations or inversions, commute with the Hamiltonian of the system. It is shown that energy and chirality eigenstates are strictly related through the Heisenberg relations, while chirality operators "conserve" parity eigenstates.

## INTRODUCTION

Chirality is the subject of everincreasing number of reports both from experimental and theorethical viewpoints in physics and chemistry.[1] This interest is easily understandable since from subatomic particles (e.g. neutrinos) to macromolecules (e.g. DNA, RNA), chirality represents a basic property of matter and the meaning of its intrinsic nature seems to be far from being fully comprehended.[2] In any case, it should be understood at a quantum mechanical level in the same way of all features and symmetries of the fundamental interactions and structures. The present article is an attempt in this direction, since we seek for a quantum mechanical treatment of central molecular chirality. Essentially, we want to show that a tetrahedral molecule, whose symmetries are described by an *O*(4)-orthogonal group, can be dealt under the standard of Schrödinger equation and Heisenberg relations of ordinary quantum mechanics.

In a recent report,[3] we proposed a geometrical representation of tetrahedral molecules starting from projections on {x,y}-plane, where the molecule center coincides with the origin of axes.

The key ingredient of such an approach is the consideration that a given bond of a molecule can be represented by a complex number, where the modulus is the "length" of the bond with respect to the stereogenic center and the "anomaly" is the angular position with respect to the other bonds and the axes. Every bond, in the {x,y}-plane, is

$$\Psi_j = \rho_j e^{i\vartheta_j} \qquad (1)$$

where $\rho_j$ is the projected length of the bond, $\vartheta_j$ is the angular position, having chosen a rotation versus; $i = \sqrt{-1}$ is the imaginary unit. A molecule with one stereogenic center is then given by the sum vector

$$M = \sum_{j=1}^{4} \rho_j e^{i\vartheta_j} \qquad (2)$$

These considerations can be immediately extended to more general cases. A molecule with *n* stereogenic centers has *n* planes of projection and the bonds between two centers have to be taken into account. A molecule with one center has four bonds, a molecule with *n* centers has $(3n+1)$ bonds. The rule works only for simply connected tetrahedrons so that a generic chain can be given by the equation

$$M_n = \sum_{k=1}^{n} \sum_{j=1}^{3n+1} \rho_{jk} e^{i\vartheta_{jk}} \qquad (3)$$

where k is the "center-index", while j is the "bond-index". A projective plane of symmetry is fixed for any k and the couple $\{\rho_{jk}, \vartheta_{jk}\} \equiv \{0,0\}$ defines the center in every plane. In another report,[4] we worked out an algebraic approach in order to deal with central molecular chirality. Starting from the empirical Fischer rules, we developed an algebra of 24 matrix operators capable of characterizing all the representations of the two enantiomers of a given tetrahedral molecule. As a result, every chain of type (3) can be classified as enantiomer, diastereoisomer or achiral molecule, thanks to the operators of such an algebra, which tell us if, with respect to a chosen fundamental representation of a given configuration for a tetrahedron, its bonds undergo a rotation or an inversion of two of their bonds.

To be more precise, let

$$M_k^{(0)} = \begin{pmatrix} \Psi_{1k} \\ \Psi_{2k} \\ \Psi_{3k} \\ \Psi_{4k} \end{pmatrix} \qquad (4)$$

be the column vector assigning the fundamental representation of a given k-tetrahedron in the chain of *n*-tetrahedrons (3). The matrix operator

$$\chi_1^k = \begin{pmatrix} 1 & 0 & 0 & 0 \\ 0 & 1 & 0 & 0 \\ 0 & 0 & 1 & 0 \\ 0 & 0 & 0 & 1 \end{pmatrix} \qquad (5)$$

leaves $M_k^{(0)}$ vector invariant, while

$$\chi_2^k = \begin{pmatrix} 0 & 0 & 1 & 0 \\ 0 & 1 & 0 & 0 \\ 0 & 0 & 0 & 1 \\ 1 & 0 & 0 & 0 \end{pmatrix} \qquad (6)$$

acts as

$$\chi_2^k \begin{pmatrix} \Psi_{1k} \\ \Psi_{2k} \\ \Psi_{3k} \\ \Psi_{4k} \end{pmatrix} = \begin{pmatrix} \Psi_{3k} \\ \Psi_{2k} \\ \Psi_{4k} \\ \Psi_{1k} \end{pmatrix} \qquad (7)$$

giving a rotation of the above fundamental representation. It is worth stressing that $\det\|\chi_2^k\|=1$ so that (6) is a rotation. On the other hand, the operator

$$\overline{\chi}_1^k = \begin{pmatrix} 0 & 0 & 0 & 1 \\ 0 & 1 & 0 & 0 \\ 0 & 0 & 1 & 0 \\ 1 & 0 & 0 & 0 \end{pmatrix} \qquad (8)$$

generates the inversion between the bonds $\Psi_{1k}$ and $\Psi_{4k}$ with respect to the fundamental representation, i.e.

$$\overline{\chi}_1^k \begin{pmatrix} \Psi_{1k} \\ \Psi_{2k} \\ \Psi_{3k} \\ \Psi_{4k} \end{pmatrix} = \begin{pmatrix} \Psi_{4k} \\ \Psi_{2k} \\ \Psi_{3k} \\ \Psi_{1k} \end{pmatrix} \qquad (9)$$

and $\det \left\| \overline{\chi}_1^k \right\| = -1$, which means that the application is an inversion. In general, the tetrahedral components of the chain (3) can be rotated or inverted in their bonds with respect to their fundamental representations. We can give the general formula

$$M_n = \sum_{k=1}^{p} \overline{M}_k + \sum_{k=p+1}^{n} M_k \qquad (10)$$

where $\overline{M}_k$ and $M_k$ are the k-tetrahedrons on which the matrix operators $\overline{\chi}_s^k$ and $\chi_s^k$ act respectively; k is the center index running from 1 to $n$; s is the operator index running from 1 to 12 (see the Appendix). The set of operators $\overline{\chi}_s^k$ and $\chi_s^k$ are elements of the algebra of $O(4)$ orthogonal group. For any k-tetrahedron in the basic configuration (4), we can have two of the 24 representations given by

$$M_k = \chi_s^k M_k^{(0)}, \qquad \overline{M}_k = \overline{\chi}_s^k M_k^{(0)} \qquad (11)$$

In this sense Eq. (10) is nothing else but Eq. (3) where the action of transformations $\chi_s^k$ and $\overline{\chi}_s^k$ is specified. Immediately, as previously discussed,[3,4] a given tetrahedral chain can be classified as

$$M_n = \sum_{k=1}^{n} M_k, \qquad p = 0 \qquad (12)$$

which is an achiral molecole on which only rotations $\chi_s^k$ act on $M_k^{(0)}$;

$$M_n = \sum_{k=1}^{p} \overline{M}_k + \sum_{k=p+1}^{n} M_k, \qquad 0 < p < n \tag{13}$$

is a diastereoisomer since $[n-(p+1)]$ tetrahedrons result superimposable after rotations, while $p$-, ones are not superimposable, having, each of them, undergone an inversion of two of their bonds. In this case, the rotations $\chi_s^k$ and the inversions $\overline{\chi}_s^k$ act on the chain $M_n$.

Finally,

$$M_n = \sum_{k=1}^{n} \overline{M}_k, \qquad n = p \tag{14}$$

is an enantiomer since the inversion operators $\overline{\chi}_s^k$ act on every tetrahedron.

In summary, central chirality is assigned by the number $\chi$ given by the couple $n, p$ that is

$$\chi \equiv \{n, p\} \tag{15}$$

where $\chi$ is the chirality index, $n$ the number of stereogenic centers and $p$ the number of inversions (at most one for each tetrahedron). The simply connected tetrahedral chains are immediately classified as $\chi \equiv \{n, 0\}$, *achiral molecules*; $\chi \equiv \{n, p < n\}$ *diastereoisomers*; and finally $\chi \equiv \{n, n\}$ *enantiomers*.

As stated above, we wonder whether such an empirical construction could be faced from a quantum mechanical point of view. In other words, we want to investigate if, in our approach, central molecular chirality can be dealt with the standard tools of quantum mechanics,[5] deriving a "quantum chiral algebra".

## THE SCHRÖDINGER EQUATION FOR A BOND IN A TETRAHEDRAL MOLECULE

Let us begin with the straightforward questions: are the bond (1) and then the sum vectors (2) and (3) solutions of a Schrödinger equation? Specifically: are there relations among chirality eigenstates, derived from the above algebra, and energy eigenstates? What is the role of parity?

In order to answer such questions, we have to discuss the possible quantum mechanical interpretation of the above formulae and the consistency of the problem.

In some reports, the dynamical treatment of linear molecular chains has been studied. For example, McLellan[6] has developed the theory of vibrational and rotational properties of a linear molecule using orthogonal finite transformations of position vectors of the atoms in the molecule. A set of generalized vector-coordinates has been identified and the invariance under the spatial transformations of the orthogonal group $O(3)$ was discussed.

King *et al.*[7] have taken into account the rotational states of a tetrahedron in a cubic crystal field. In this case, the Schrödinger equation is solved for a rigid cube, whose center of mass is fixed at a point of symmetry in an external field. Quantum states are classified under the direct product group $\overline{O} \otimes O$, where $\overline{O}$ is the octahedral group of rotations about body-fixed axes and $O$ is the group of rotations about space-fixed axes. The Born-Oppenheimer approximation is used in both situations.

In our case, we will proceed by an inverse problem approach, as usual in several branches of mathematical physics.[8] Our aim is to "construct" a suitable Schrödinger equation for the "solution" (1) and the sums (2) and (3).

First of all, the problem can be set in the Born-Oppenheimer approximation: this means that a given stereogenic center is considered fixed and dynamics of the four bonds is reduced to it.

An obvious objection to this position could be the fact that not always the center of mass of the tetrahedron coincides with the stereogenic center. However, in any case, by an appropriate change of coordinates, our problem can be reduced to a coordinate system fixed in the stereogenic center. With this position in mind, let us consider a time-dependent three-dimensional Schrödinger equation

$$i\hbar \frac{\partial \Phi}{\partial t} = \hat{H}\, \Phi, \tag{16}$$

where $\Phi = \Phi(\vec{r}, t)$ is a function of radius $\vec{r}$ and time t. The form of Hamiltonian operator specifies the problem. We assume the separation of variables since, as standard, the potential in $\hat{H}$ is independent of time, so that the problem reduces to

$$\Phi(\vec{r}, t) = \eta(\vec{r}) e^{-\frac{iE_j t}{\hbar}}, \qquad \hat{H}\eta(\vec{r}) = E_j\, \eta(\vec{r}) \tag{17}$$

where $E_j$ are the energy eigenvalues. The index j will be defined below. Explicitly, we have

$$\left\{-\frac{\hbar^2}{2\mu}\nabla^2 + V(x,y,z)\right\}\eta(x,y,z) = E_j\,\eta(x,y,z) \tag{18}$$

where "$\mu$" is a given reduced mass and $\nabla^2$ the Laplace operator. Let us pass now to a system of polar coordinates $\{r,\vartheta,\varphi\}$ by the transformations

$$x = r\sin\varphi\cos\vartheta, \quad y = r\sin\varphi\sin\vartheta, \quad z = r\cos\varphi \tag{19}$$

Our tetrahedron can be idealized as a system in a central potential with a spherical symmetry. In this case, the potential depends only on the radius r, so that $V(\vec{r}) \equiv V(r)$. By expressing the Laplace operator in terms of transformations (19), we get

$$\left[\frac{1}{r^2}\frac{\partial}{\partial r}\left(r^2\frac{\partial}{\partial r}\right) + \frac{1}{r^2\sin\varphi}\frac{\partial}{\partial\varphi}\left(\sin\varphi\frac{\partial}{\partial\varphi}\right) + \frac{1}{r^2\sin^2\varphi}\frac{\partial^2}{\partial\vartheta^2} - U(r) + \frac{2\mu E_j}{\hbar^2}\right]\eta(r,\vartheta,\varphi) = 0 \tag{20}$$

where $U(r) = \frac{2\mu}{\hbar^2}V(r)$. Again, the problem can be separated by assuming

$$\eta(r,\vartheta,\varphi) = Y(\vartheta,\varphi)\rho(r) \tag{21}$$

and then

$$\frac{1}{\rho}\frac{d}{dr}\left(r^2\frac{d\rho}{dr}\right) + \left[\frac{2\mu E_j}{\hbar^2} - U(r)\right]r^2 = \lambda \tag{22}$$

$$\frac{1}{Y}\frac{1}{\sin\varphi}\frac{\partial}{\partial\varphi}\left(\sin\varphi\frac{\partial Y}{\partial\varphi}\right) + \frac{1}{Y}\frac{1}{\sin^2\varphi}\frac{\partial^2 Y}{\partial\vartheta^2} = -\lambda \tag{23}$$

The problem can be furtherly separated introducing the angular momentum operator and its projection along the z-axis,

$$L^2 = -\hbar^2\left[\frac{1}{\sin\varphi}\frac{\partial}{\partial\varphi}\left(\sin\varphi\frac{\partial}{\partial\varphi}\right) + \frac{1}{\sin^2\varphi}\frac{\partial^2}{\partial\vartheta^2}\right] \tag{24}$$

$$L_z = \frac{\hbar}{i}\frac{\partial}{\partial\vartheta} \tag{25}$$

and then

$$L^2 Y(\vartheta,\varphi) = \hbar^2 l(l+1) Y(\vartheta,\varphi) \tag{26}$$

$$L_z Y(\vartheta,\varphi) = m\hbar Y(\vartheta,\varphi) \tag{27}$$

with the conditions

$$\lambda = l(l+1) \qquad l \geq |m| \tag{28}$$

The general solution of the angular part of Schrödinger equation is then

$$Y_{l,m}(\vartheta,\varphi) = N_{l,m} P_l^m(\cos\varphi)(\sin\varphi)^{|m|} e^{im\vartheta} \tag{29}$$

where $P_l^m(\cos\varphi)$ are Legendre polynomials and

$$N_{l,m} = \left[\frac{(2l+1)(l-|m|)!}{2(l+|m|)!} \frac{1}{2\pi}\right]^{\frac{1}{2}} \tag{30}$$

is a normalization factor depending on $l$ and $m$. Let us now return to our former problem. We want to see if Eq. (1) is a solution of Schrödinger equation. Concerning the angular component, it can be interpreted as the azimuthal part of angular momentum.

For the radial component, let us assume, for the sake of simplicity,

$$\rho(r) = \left(\frac{r}{r_0}\right) \tag{31}$$

where $r_0$ is a normalization length (e.g. ≈ 1.09÷1.54Å a typical C—X length bond) useful to restore the probabilistic interpretation of our approach. Immediately, we obtain the form of the potential $V(r)$, which is

$$V(r) = E_j - \frac{\alpha_0}{r^2}, \qquad \alpha_0 = \frac{\hbar^2}{2\mu}[l(l+1) - 2] \tag{32}$$

depending on the eigenvalue $E_j$, the angular momentum $l$ and the mass $\mu$.

In conclusion, the equation $\Psi j = \rho_j(r) e^{i\vartheta_j}$ with the positions

$$\rho_j(r) = \left(\frac{r_j}{r_0}\right), \qquad \vartheta_j = m_j \vartheta \tag{33}$$

can be considered as the radial and azimuthal components of a solution of Schrödinger equation. Clearly, j = 1,2,3,4, $E_j$ are the bonds energies and $M = \sum_{j=1}^{4} \rho_j e^{i\vartheta_j}$ is the superposition of four single particular solutions. This result can be easily extended to Eq. (3) considering *n* Schrödinger problems, one for each stereogenic center. However, boundary conditions between two connected stereogenic centers have to be chosen carefully. Some comments are necessary at this point. Clearly, the full quantum problem of tetrahedral molecules has not been taken into account. For example, we have not given a complete treatment of vibrational and rotational states of the molecule. However, this is not our goal, since we want just to focus the attention on the fact that the representation and algebra previously developed[3,4] can have an interpretation at a fundamental level. Besides, we are going to investigate if the operators $\chi_s^k$ and $\overline{\chi}_s^k$ can be read as quantum operators acting on the bonds of a given tetrahedral molecule.

## CENTRAL MOLECULAR CHIRALITY AS A QUANTUM ALGEBRA

The above treatment shows that Eq. (1) can be considered as a solution of a "reduced" Schrödinger problem, where, due to the separation of variables, the Hamiltonian operator is projected on the $\{r,\vartheta\}$- plane and a part of the general solution $\Phi = \Phi(r,\vartheta,\varphi,t)$ is nothing else but $\Psi(r,\vartheta) = \rho e^{i\vartheta}$. Let now us take into account the operators $\hat{H}_k(r,\vartheta)$, $\chi_s^k$ and $\overline{\chi}_s^k$ which act on the four bonds of the $k^{th}$-stereogenic center. They induce on the tetrahedron the following transformations

$$i\hbar \frac{\partial}{\partial t} M_k = \hat{H}_k M_k, \qquad M_k = \chi_s^k M_k^{(0)} \qquad (34)$$

and

$$i\hbar \frac{\partial}{\partial t} \overline{M}_k = \hat{H}_k \overline{M}_k, \qquad \overline{M}_k = \overline{\chi}_s^k M_k^{(0)} \qquad (35)$$

It is clear that the rotations $\chi_s^k$ and the inversions $\overline{\chi}_s^k$ operate on the bonds of the starting fundamental representation $M_k^{(0)}$ given by Eq. (4). Then, we can write

$$i\hbar \frac{\partial}{\partial t} M_k = i\hbar \frac{\partial}{\partial t} \left( \chi_s^k M_k^{(0)} \right) = \hat{H}_k \chi_s^k M_k^{(0)} \tag{36}$$

On the other hand, we have

$$\chi_s^k \left( i\hbar \frac{\partial}{\partial t} M_k^{(0)} \right) = i\hbar \frac{\partial}{\partial t} \left( \chi_s^k M_k^{(0)} \right) = \chi_s^k \hat{H}_k M_k^{(0)} \tag{37}$$

so that subtracting Eq. (37) from Eq. (36) we obtain

$$\left( \hat{H}_k \chi_s^k - \chi_s^k \hat{H}_k \right) M_k^{(0)} = \left[ \hat{H}_k, \chi_s^k \right] M_k^{(0)} = 0 \tag{38}$$

that is $\chi_s^k$ and $\hat{H}_k$ commute between them being

$$\left[ \hat{H}_k, \chi_s^k \right] = 0 \tag{39}$$

Similarly, we obtain

$$\left[ \hat{H}_k, \overline{\chi}_s^k \right] = 0 \tag{40}$$

Furthermore (see also the Appendix and ref. 4), we have

$$\left[ \chi_s^k, \chi_m^k \right] = 0 \quad \text{for } s, m = 5, 7, 11 \tag{41}$$

$$\left[ \overline{\chi}_s^k, \overline{\chi}_m^k \right] = \chi_l - \chi_j \quad \text{for } l, m, j, s = 1, \ldots, 12 \tag{42}$$

$$\left[ \chi_s^k, \chi_m^k \right] = \chi_l - \chi_j \quad \text{for } l, m, j, s = 1, \ldots, 12 \text{ with } s, m \neq 5, 7, 11 \tag{43}$$

$$\left[ \chi_s^k, \overline{\chi}_m^k \right] = \overline{\chi}_l - \overline{\chi}_j \quad \text{for } l, m, j, s = 1, \ldots, 12 \tag{44}$$

and finally

$$\left[ \chi_s^k, \chi_m^j \right] = 0 \tag{45}$$

$$\left[ \overline{\chi}_s^k, \overline{\chi}_m^j \right] = 0 \tag{46}$$

$$\left[ \overline{\chi}_s^k, \chi_m^j \right] = 0 \tag{47}$$

being $k \neq j$, as it is obvious for different tetrahedrons.

As final remark, it can be said that the relations (39)-(47) constitute a quantum chiral algebra and the eigenstates of $\chi_s^k$ and $\bar{\chi}_s^k$ operators are, in general, solutions of Schrödinger equation.

## QUANTUM CHIRAL ALGEBRA AND PARITY

As previously seen, the set of operators $\hat{H}_k$, $\chi_s^k$, $\bar{\chi}_s^k$ for a given k-tetrahedron in the s-representations give rise to an algebra. Now we want to discuss how these operators effectively act on quantum states of chiral molecules and how the results can be interpreted.

Let us take into account the fundamental representation (4) of a given k-tetrahedron. The action of the operator $\chi_s^k$ and $\bar{\chi}_s^k$ defines the "chiral state" of the molecule being, as we have seen,

$$|\Psi_R\rangle = M_k = \chi_s^k M_k^{(0)} \qquad (48)$$

and

$$|\Psi_L\rangle = \overline{M}_k = \bar{\chi}_s^k M_k^{(0)} \qquad (49)$$

where we are indicating with $|\Psi_R\rangle$ and $|\Psi_L\rangle$ the right- and left-handed quantum states of the molecule using the Dirac Ket notation. Clearly the operators $\chi_s^k$ "rotate" the k-tetrahedron, while $\bar{\chi}_s^k$ "invert" a couple of bonds. Dropping, for simplicity, the indexes, we have the relations

$$\chi|\Psi_R\rangle = |\Psi_R\rangle \ ; \ \chi|\Psi_L\rangle = |\Psi_L\rangle \qquad (50)$$

$$\bar{\chi}|\Psi_R\rangle = |\Psi_L\rangle \ ; \ \bar{\chi}|\Psi_L\rangle = |\Psi_R\rangle \qquad (51)$$

where the $\bar{\chi}$ operators interconvert the two handed forms and, in some cases, act as an algebraic counterpart of quantum tunnelling.[2a,9] Besides, the parity eigenstates of a chiral molecule, ignoring eventual parity violation effects,[10] are energy eigenstates and can be obtained as superpositions of handed states.[11] We have

$$|\Psi_\pm\rangle = \frac{1}{\sqrt{2}}(|\Psi_L\rangle \pm |\Psi_R\rangle) \qquad (52)$$

which are, respectively, even- and odd- parity eigenstates. The chirality operators $\chi$ and $\bar{\chi}$ do not change the parity of a given enantiomer being

$$\chi|\Psi_\pm\rangle = \frac{1}{\sqrt{2}}\left(\chi|\Psi_L\rangle \pm \chi|\Psi_R\rangle\right) \tag{53}$$

$$= \frac{1}{\sqrt{2}}\left(|\Psi_L\rangle \pm |\Psi_R\rangle\right)$$

and

$$\bar{\chi}|\Psi_\pm\rangle = \frac{1}{\sqrt{2}}\left(\bar{\chi}|\Psi_L\rangle \pm \bar{\chi}|\Psi_R\rangle\right) \tag{54}$$

$$= \frac{1}{\sqrt{2}}\left(|\Psi_R\rangle \pm |\Psi_L\rangle\right)$$

In summary, the $\bar{\chi}$-operators allow the transition among the $|\Psi_R\rangle$-state, the $|\Psi_L\rangle$-state and viceversa (as a quantum tunnelling process), but parity is a conserved quantum mechanical quantity in the Hund sense.[12]

An important remark has to be done at this point. The total Hamiltonian operator for the degenerate isomers of an optically molecule always consists of an even and an odd part.[13]

$$\hat{H}^{tot} = \hat{H}^{even} + \hat{H}^{odd} \tag{55}$$

This is the energy operator involved in the Hund result

$$\left[\hat{P}, \hat{H}^{tot}\right] = 0 \tag{56}$$

On the other hand, the Hamiltonian operators which we have considered above (i.e. $\hat{H}_k$) refer to the construction in which we studied the bond eigenstates related to the k-tetrahedron chiral eigenstates. Due to relations (39) and (40), these eigenstates are "conserved" with respect to the Hamiltonian $\hat{H}_k$ and not with respect to the total Hamiltonian $\hat{H}^{tot}$. With these considerations in mind, parity is the true conserved quantum-mechanical quantity and not chirality. In a forthcoming paper the effects of parity violation with respect to the proposed quantum chiral algebra will be discussed.

## ACKNOWLEDGEMENTS

The authors wish to thank a referee for the useful suggestions and comments which allowed to improve the reported results.

## Appendix

Here we report the 24 matrix-transformations deduced from the Fischer projections (12 for (+) enantiomer and 12 for (−) enantiomer of a generic chiral tetrahedral molecule with one stereogenic center).[4] The following charts summarize the situation. The operators $\chi_s$ give rise to the representations of the (+) enantiomer, while the operators $\bar{\chi}_s$ give rise to those of the (−) enantiomer starting from a basic representation. Obviously $s = 1,..,12$. The matrices in Chart I and II are the elements of a 4-parameter algebra. Those in Chart I are a representation of rotations, while those in Chart II are inversions. Both sets constitute the $O(4)$ group of $4 \times 4$ orthogonal matrices.

Chart I, (+)-enantiomer:

$$\chi_1 = \begin{pmatrix} 1 & 0 & 0 & 0 \\ 0 & 1 & 0 & 0 \\ 0 & 0 & 1 & 0 \\ 0 & 0 & 0 & 1 \end{pmatrix} \quad \chi_2 = \begin{pmatrix} 0 & 0 & 1 & 0 \\ 0 & 1 & 0 & 0 \\ 0 & 0 & 0 & 1 \\ 1 & 0 & 0 & 0 \end{pmatrix} \quad \chi_3 = \begin{pmatrix} 0 & 0 & 0 & 1 \\ 0 & 1 & 0 & 0 \\ 1 & 0 & 0 & 0 \\ 0 & 0 & 1 & 0 \end{pmatrix}$$

$$\chi_4 = \begin{pmatrix} 0 & 1 & 0 & 0 \\ 0 & 0 & 1 & 0 \\ 1 & 0 & 0 & 0 \\ 0 & 0 & 0 & 1 \end{pmatrix} \quad \chi_5 = \begin{pmatrix} 0 & 1 & 0 & 0 \\ 1 & 0 & 0 & 0 \\ 0 & 0 & 0 & 1 \\ 0 & 0 & 1 & 0 \end{pmatrix} \quad \chi_6 = \begin{pmatrix} 0 & 1 & 0 & 0 \\ 0 & 0 & 0 & 1 \\ 0 & 0 & 1 & 0 \\ 1 & 0 & 0 & 0 \end{pmatrix}$$

$$\chi_7 = \begin{pmatrix} 0 & 0 & 1 & 0 \\ 0 & 0 & 0 & 1 \\ 1 & 0 & 0 & 0 \\ 0 & 1 & 0 & 0 \end{pmatrix} \quad \chi_8 = \begin{pmatrix} 0 & 0 & 0 & 1 \\ 1 & 0 & 0 & 0 \\ 0 & 0 & 1 & 0 \\ 0 & 1 & 0 & 0 \end{pmatrix} \quad \chi_9 = \begin{pmatrix} 1 & 0 & 0 & 0 \\ 0 & 0 & 1 & 0 \\ 0 & 0 & 0 & 1 \\ 0 & 1 & 0 & 0 \end{pmatrix}$$

$$\chi_{10} = \begin{pmatrix} 1 & 0 & 0 & 0 \\ 0 & 0 & 0 & 1 \\ 0 & 1 & 0 & 0 \\ 0 & 0 & 1 & 0 \end{pmatrix} \quad \chi_{11} = \begin{pmatrix} 0 & 0 & 0 & 1 \\ 0 & 0 & 1 & 0 \\ 0 & 1 & 0 & 0 \\ 1 & 0 & 0 & 0 \end{pmatrix} \quad \chi_{12} = \begin{pmatrix} 0 & 0 & 1 & 0 \\ 1 & 0 & 0 & 0 \\ 0 & 1 & 0 & 0 \\ 0 & 0 & 0 & 1 \end{pmatrix}$$

Chart II, (−) enantiomer:

$$\bar{\chi}_1 = \begin{pmatrix} 0 & 0 & 0 & 1 \\ 0 & 1 & 0 & 0 \\ 0 & 0 & 1 & 0 \\ 1 & 0 & 0 & 0 \end{pmatrix} \quad \bar{\chi}_2 = \begin{pmatrix} 0 & 0 & 1 & 0 \\ 0 & 1 & 0 & 0 \\ 1 & 0 & 0 & 0 \\ 0 & 0 & 0 & 1 \end{pmatrix} \quad \bar{\chi}_3 = \begin{pmatrix} 1 & 0 & 0 & 0 \\ 0 & 1 & 0 & 0 \\ 0 & 0 & 0 & 1 \\ 0 & 0 & 1 & 0 \end{pmatrix}$$

$$\bar{\chi}_4 = \begin{pmatrix} 0 & 1 & 0 & 0 \\ 0 & 0 & 1 & 0 \\ 0 & 0 & 0 & 1 \\ 1 & 0 & 0 & 0 \end{pmatrix} \quad \bar{\chi}_5 = \begin{pmatrix} 0 & 1 & 0 & 0 \\ 0 & 0 & 0 & 1 \\ 1 & 0 & 0 & 0 \\ 0 & 0 & 1 & 0 \end{pmatrix} \quad \bar{\chi}_6 = \begin{pmatrix} 0 & 1 & 0 & 0 \\ 1 & 0 & 0 & 0 \\ 0 & 0 & 1 & 0 \\ 0 & 0 & 0 & 1 \end{pmatrix}$$

$$\bar{\chi}_7 = \begin{pmatrix} 0 & 0 & 1 & 0 \\ 1 & 0 & 0 & 0 \\ 0 & 0 & 0 & 1 \\ 0 & 1 & 0 & 0 \end{pmatrix} \quad \bar{\chi}_8 = \begin{pmatrix} 1 & 0 & 0 & 0 \\ 0 & 0 & 0 & 1 \\ 0 & 0 & 1 & 0 \\ 0 & 1 & 0 & 0 \end{pmatrix} \quad \bar{\chi}_9 = \begin{pmatrix} 0 & 0 & 0 & 1 \\ 0 & 0 & 1 & 0 \\ 1 & 0 & 0 & 0 \\ 0 & 1 & 0 & 0 \end{pmatrix}$$

$$\bar{\chi}_{10} = \begin{pmatrix} 0 & 0 & 0 & 1 \\ 1 & 0 & 0 & 0 \\ 0 & 1 & 0 & 0 \\ 0 & 0 & 1 & 0 \end{pmatrix} \quad \bar{\chi}_{11} = \begin{pmatrix} 1 & 0 & 0 & 0 \\ 0 & 0 & 1 & 0 \\ 0 & 1 & 0 & 0 \\ 0 & 0 & 0 & 1 \end{pmatrix} \quad \bar{\chi}_{12} = \begin{pmatrix} 0 & 0 & 1 & 0 \\ 0 & 0 & 0 & 1 \\ 0 & 1 & 0 & 0 \\ 1 & 0 & 0 & 0 \end{pmatrix}$$